\RequirePackage{lineno}

\documentclass[aps,prd,twocolumn,groupedaddress,nofootinbib,showpacs]{revtex4}%

\usepackage{graphicx}

\usepackage{amsmath}
\usepackage{amssymb}
\usepackage{multirow}
\usepackage{bm}
\usepackage{color}

\usepackage[hypertex]{hyperref}

\def\be{\begin{linenomath*}\begin{equation}}
\def\ee{\end{equation}\end{linenomath*}}
\def\ba{\begin{linenomath*}\begin{eqnarray}}
\def\ea{\end{eqnarray}\end{linenomath*}}
\def\ss{{\bigl.^3\hspace{-1mm}S^{[1]}_1}}

\def\p0{{\bigl.^3\hspace{-1mm}P^{[8]}_0}}

\def\s{\sigma}

\begin{document}



\title{\Large $\Upsilon(nS)$ and $\chi_b(nP)$ production at hadron colliders in nonrelativistic QCD }
\author{
 Hao Han$^a$, Yan-Qing Ma$^{a,b,c}$, Ce Meng$^a$, Hua-Sheng Shao$^a$, Yu-Jie Zhang$^d$, Kuang-Ta Chao$^{a,c,e}$}

\affiliation{ {\footnotesize (a)~School of Physics and State Key
Laboratory of Nuclear Physics and Technology, Peking University,
Beijing 100871, China}\\
{\footnotesize (b)~Maryland Center for Fundamental Physics, University of Maryland, College Park, Maryland 20742, USA}\\
 {\footnotesize (c)~Center for High Energy physics, Peking
University, Beijing 100871, China}\\
{\footnotesize (d)~Key Laboratory of Micro-nano
 Measurement-Manipulation and Physics (Ministry of Education) and School of Physics, Beihang University, Beijing 100191, China}\\
 {\footnotesize (e)Collaborative Innovation Center of Quantum Matter, Beijing 100871, China}
}

\begin{abstract}
$\Upsilon(nS)$ and $\chi_b(nP)$ (n=1,2,3) production at the LHC is studied at  next-to-leading order in $\alpha_s$ in nonrelativistic QCD. Feeddown contributions from higher $\chi_b$ and $\Upsilon$ states are all considered for lower $\Upsilon$ cross sections and polarizations. The long distance matrix elements (LDMEs) are extracted from the yield data, and then used to make
predictions for the $\Upsilon(nS)$  polarizations, which are found to be consistent with the measured polarization data within errors. In particular, the $\Upsilon(3S)$ polarization puzzle can be understood by a large feeddown contribution from $\chi_b(3P)$ states. Our results may provide a good description for both cross sections and polarizations of prompt $\Upsilon(nS)$ and $\chi_b(nP)$ production at the LHC.

\end{abstract}


\maketitle


\section{Introduction}

Since the surprisingly large production rate of $\psi'$ at large $p_T$ was found by CDF in 1992~\cite{Abe:1992ww}, the production of heavy quarkonium at hadron colliders has been a problem full of puzzles. While the color-octet (CO) mechanism~\cite{Braaten:1994vv} at leading order (LO) in nonrelativistic QCD (NRQCD) factorization~\cite{Bodwin:1994jh} might explain the large production rates of $\psi'$ and $J/\psi$ at large $p_T$ via gluon fragmentation, the predicted transverse polarizations for $J/\psi(\psi')$ were in contradiction with the measurements that the produced $J/\psi(\psi')$ were almost unpolarized (see Ref.~\cite{Brambilla:2010cs} for a comprehensive review). In recent years, significant progress has been made in the next-to-leading order (NLO) QCD calculations in NRQCD. Calculations and fits for both yield and polarization in $J/\psi$ production are performed by three groups~\cite{Butenschoen:2012px,Chao:2012iv,Gong:2012ug}, but the conclusions are quite different. In Ref.\cite{Chao:2012iv} a simultaneous description for the observed $J/\psi$ yield and polarization can be achieved at large $p_T$ ($>$7~GeV) by considering possible cancelations between contributions of S- and P-wave color-octet channels. Recently, by including leading power fragmentation corrections, which improves the convergence of $\alpha_s$ expansion at large $p_T$, a good explanation for the $J/\psi$ polarization
is also found\cite{Bodwin:2014gia}.

Recently, polarizations of $\Upsilon(1S,2S,3S)$ have been measured by CMS at the LHC~\cite{Chatrchyan:2012woa}. It is interesting to study the $\Upsilon$ production within the same framework as that for the $J/\psi$ production and further test the interpretation for the polarization puzzle in Ref.\cite{Chao:2012iv}. Note that $\Upsilon$ should be a more suitable system than $J/\psi$ to apply NRQCD, since both $v$ (the relative velocity of heavy quarks in heavy quarkonium) and $\alpha_s$
are smaller for bottomonium than charmonium, and thus the double expansion in $\alpha_s$ and $v$ should converge faster for bottomonium production. Earlier studies of $\Upsilon$ and $\chi_b$ production can be found in Refs.~\cite{Braaten:2000gw,Artoisenet:2008fc,Wang:2012is,Likhoded:2012hw} and references therein. In Ref.\cite{Gong:2013qka},
a NLO calculation of $\Upsilon(1S,2S,3S)$ polarizations is given, where the polarizations for $\Upsilon(1S,2S)$ agree with the CMS measurements~\cite{Chatrchyan:2012woa}, but the predicted ratio of differential cross sections of $\chi_{b2}(1P)$ to $\chi_{b1}(1P)$\cite{Gong:2013qka} is too large and inconsistent with the CMS data\cite{Khachatryan:2014ofa}. Furthermore, without considering the $\chi_b(3P)$ feeddown, the polarization data of $\Upsilon(3S)$ can not be explained \cite{Gong:2013qka}.


Recently, the radiative transition of $\chi_b(3P)$ to $\Upsilon(3S)$ was first seen by LHCb~\cite{Aaij:2014caa}. So the explanation of $\Upsilon(1S,2S)$ and $\Upsilon(3S)$ polarizations should be reconsidered, and a proper treatment for $\chi_b(1P,2P,3P)$ feeddown is needed, since the treatment of  $\chi_b(3P)$ and $\Upsilon(3S)$ will affect the production of $\Upsilon(1S,2S)$ through the cascaded effects.
In this work, we study the prompt production of $\Upsilon(1S,2S,3S)$ with both  direct and feeddown contributions at NLO in $\alpha_s$ in NRQCD.

The polarized cross section for a bottomonium $H$ can be factorized as~\cite{Bodwin:1994jh}
\begin{eqnarray}\label{Factorization-NRQCD}
&d\sigma_{s_z,s_z}=\sum_{i,j,n} \int dx_1dx_2\, G_{i/p}G_{j/p}
\langle\mathcal{O}^H_n\rangle\, d\hat{\sigma}_{s_z,s_z}^{i,j,n},
\end{eqnarray}
where $p$ denotes either proton or anti-proton, $G_{i,j/p}$ are the parton distribution functions (PDFs) of $p$, and the indices $i, j$ run over all
the partonic species. $\langle\mathcal{O}^H_n\rangle$ is the long distance matrix element (LDME), with ``$n$'' denotes the color, spin and angular momentum of the intermediate $b\bar{b}$ pair, which can be $^3\!S_1^{[1,8]}$, $^1\!S_0^{[8]} $ and $^3\!P_J^{[8]}$ for $\Upsilon$, and $^3\!P_J^{[1]}$ and $^3\!S_1^{[8]} $ for $\chi_b$. The yield can be obtained by summing the polarized cross sections over the spin quantum number $s_z$. The virtual corrections are calculated by using our Mathematica code~\cite{Ma:2010yw,Ma:2010jj,Chao:2012iv}, and the real corrections are obtained by using the HELAC-Onia program~\cite{Shao:2012iz}.
We further use the CTEQ6L1 and CTEQ6M PDFs~\cite{Whalley:2005nh} respectively for LO and NLO calculations. The bottom quark mass is set to be $m_b = 4.75$~GeV,  the renormalization, factorization, and NRQCD scales are $\mu_r = \mu_f = \sqrt{p_T^2+4m_b^2}$ and $\mu_{\Lambda} = m_b$.


\section{Feeddown and $\chi_b(nP)$}

For $\Upsilon$ the polarization observable $\lambda_\theta$  can be expressed as $\lambda_{\theta}=\frac{\rm{d}\sigma_{11}-\rm{d}\sigma_{00}}{\rm{d}\sigma_{11}+\rm{d}\sigma_{00}}$,
where $\sigma_{00}$ and $\sigma_{11}$ are polarized prompt cross sections, including both direct production
and feeddown contributions from higher $\Upsilon(nS)$ and $\chi_b(nP)$ states. Since the transitions between $\Upsilon(nS)$ are dominated by the S-wave dipion modes, the feeddown of higher $\Upsilon(nS)$ will inherit the spin index of the mother particles. While for the $\chi_b(nP)$ feeddown, which proceeds mainly through $\chi_b(nP)\to\Upsilon(mS)\gamma$, the general inheritance relations of polarizations are given in Refs.~\cite{Shao:2012fs,Shao:2014fca}:
\begin{eqnarray}\label{lamda0}
\lambda_{\theta}^{\chi_{b0}\to\Upsilon} &=& 0,\nonumber\\
\label{lamda-chib-feeddown}\lambda_{\theta}^{\chi_{b1}\to\Upsilon} &=& \frac{\rm{d}\sigma_{00}^{\chi_{b1}}-\rm{d}\sigma_{11}^{\chi_{b1}}}{3\rm{d}\sigma_{11}^{\chi_{b1}}+\rm{d}\sigma_{00}^{\chi_{b1}}},\\
\lambda_{\theta}^{\chi_{b2}\to\Upsilon} &=& \frac{6\rm{d}\sigma_{22}^{\chi_{b2}}-3\rm{d}\sigma_{11}^{\chi_{b2}}-3\rm{d}\sigma_{00}^{\chi_{b2}}} {6\rm{d}\sigma_{22}^{\chi_{b2}}+9\rm{d}\sigma_{11}^{\chi_{b2}}+5\rm{d}\sigma_{00}^{\chi_{b2}}}\nonumber.
\end{eqnarray}
Similar to $\chi_{cJ}$\cite{Ma:2010vd}, at NLO in $\alpha_s$ the $\chi_{bJ}$ production is determined by the color-octet (CO) $^3\!S_1^{[8]}$ and color-singlet (CS) $^3\!P_J^{[1]}$ contributions. If CO $^3\!S_1^{[8]}$ is dominant, which leads to transverse polarization at large $p_T$, the ratios of polarized cross sections become $\rm{d}\sigma_{00}^{\chi_{b1}}:\rm{d}\sigma_{11}^{\chi_{b1}} = 2:1$ and $\rm{d}\sigma_{00}^{\chi_{b2}}:\rm{d}\sigma_{11}^{\chi_{b2}}:\rm{d}\sigma_{22}^{\chi_{b2}} = 1/3:1/2:1$, and the feeddown polarization parameters in Eq.~(\ref{lamda-chib-feeddown}) are 0.20 for $\chi_{b1}$ and 0.29 for $\chi_{b2}$. Further including the CS $^3\!P_J^{[1]}$ contribution only slightly changes the  overall
polarization of $\chi_{bJ}$ feeddown.
This shows that the $\chi_b$ feeddown contributes a modest transverse polarization for $\Upsilon$ at large $p_T$.

The CS LDMEs for $\chi_{bJ}(nP)$ can be related to the derivatives of radial wave functions at the origin by
\be\label{3PJ1}
\langle\mathcal{O}^{\chi_{bJ}(nP)}(^3\!P_J^{[1]})\rangle=(2J+1)\frac{3}{4\pi}|R^{\prime}_{nP}(0)|^2,
\ee
where $|R^{\prime}_{nP}(0)|^2$  can be estimated in potential models. E. g. the B-T potential model\cite{Eichten:1995ch} gives $|R^{\prime}_{1P,2P,3P}(0)|^2=(1.417,\ 1.653,\ 1.794)\ \mbox{GeV}^5$. In fact, various potentials in Refs.\cite{Eichten:1995ch} and \cite{Kwong:1988ae} all indicate $|R^{\prime}_{1P}(0)|^2\approx|R^{\prime}_{2P}(0)|^2\approx|R^{\prime}_{3P}(0)|^2$. So, as a balanced approximation, we use
\be\label{R0}
|R^{\prime}_{nP}(0)|^2 \approx 1.653\ \mbox{GeV}^5,\ \ \ n=1, 2, 3,
\ee
as input. The CO LDMEs are introduced via the ratio
\be\label{r-nP}
r_{nP}=m_b^2\langle\mathcal{O}^{\chi_{bJ}(nP)}(^3\!S_1^{[8]}) \rangle/\langle\mathcal{O}^{\chi_{bJ}(nP)}(^3\!P_J^{[1]})\rangle,
\ee
which is independent of $J$ since $\langle\mathcal{O}^{\chi_{bJ}(nP)}(^3\!S_1^{[8]})\rangle=(2J+1)\langle\mathcal{O}^{\chi_{b0}(nP)}(^3\!S_1^{[8]})\rangle$.
Unlike the CS LDMEs, $r_{nP}$ can not be estimated from potential models, but should be extracted from experimental data.

\begin{table}[!htp]
\begin{tabular}{{{c}cc}}
\hline\hline \itshape
~$Br$~&~$theory$~&~$Experiment$\cite{Beringer:1900zz}~
\\\hline $\chi_{b1}(2P)\rightarrow\Upsilon(2S)$ & $15.6\%$ & $19.9\pm 1.9\%$
\\\hline $\chi_{b1}(2P)\rightarrow\Upsilon(1S)$ & $9.7\%$ & $9.2\pm 0.8\%$
\\\hline $\chi_{b2}(2P)\rightarrow\Upsilon(2S)$ & $8.3\%$ & $10.6\pm 2.6\%$
\\\hline $\chi_{b2}(2P)\rightarrow\Upsilon(1S)$ & $7.3\%$ & $7.0\pm 0.7\%$
\\\hline\hline
\end{tabular}
\caption{\label{br0} Predicted branching ratios $\mbox{Br}(\chi_{b1,b2}(2P)\rightarrow\Upsilon(1S,2S)\gamma)$ by assuming the total decay widths of $\chi_{bJ}(nP)$ are independent of $n$, as compared with experiments\cite{Beringer:1900zz}.}
\end{table}
\begin{table}[!htp]
\begin{tabular}{{{c}ccc}}
\hline\hline \itshape
~$Br$~&~$n = 1$~&~$n = 2$~&~$n = 3$~
\\\hline $\chi_{b0}(3P)\rightarrow\Upsilon(nS)$ & $0.24\%$ & $0.22\%$ & $0.50\%$
\\\hline $\chi_{b1}(3P)\rightarrow\Upsilon(nS)$ & $3.81\%$ & $3.68\%$ & $10.44\%$
\\\hline $\chi_{b2}(3P)\rightarrow\Upsilon(nS)$ & $1.92\%$ & $1.91\%$ & $6.11\%$
\\\hline\hline
\end{tabular}
\caption{\label{br} Predicted branching ratios $\mbox{Br}(\chi_{bJ}(3P)\rightarrow\Upsilon(1S,2S,3S)\gamma)$ by assuming the total decay widths of $\chi_{bJ}(nP)$ are independent of $n$.}
\end{table}

 We also assume that the total decay widths of $\chi_{bJ}(nP)$, which are related to $|R^{\prime}_{nP}(0)|^2$, are approximately independent of $n$. Then, taking the partial decay widths of $\chi_{bJ}(nP)\rightarrow\Upsilon(mS)\gamma$ calculated in Ref.\cite{Kwong:1988ae} and the PDG values of $\mbox{Br}(\chi_{bJ}(1P)\rightarrow\Upsilon(1S)\gamma)$~\cite{Beringer:1900zz} as inputs, we can calculate the branching ratios $\mbox{Br}(\chi_{bJ}(2P)\rightarrow\Upsilon(2S)\gamma)$ and $\mbox{Br}(\chi_{bJ}(2P)\rightarrow\Upsilon(1S)\gamma)$, which are found to be close to their PDG values\cite{Beringer:1900zz}, as shown in Tab.~\ref{br0}.   This implies that it may be a good approximation that
 the total widths of $\chi_b(nP)$ are independent of $n$. The above approximation is also roughly consistent with the recent calculations based on the potential model in~\cite{Godfrey:2015dia}. With this approximation we further calculate $\mbox{Br}(\chi_{bJ}(3P)\rightarrow\Upsilon(1S,2S,3S)\gamma)$, which are listed in Tab.~\ref{br}.


\section{Prompt $\Upsilon(nS)$ production}

Having clarified how to treat the feeddown contributions, we now extract LDMEs of $\Upsilon(nS)$ and $r_{nP}$ defined in (\ref{r-nP}) by fitting the yield data at LHC, and leave polarizations as our prediction. Data in our fit includes: (1) Differential cross sections of $\Upsilon(nS)$ measured by ATLAS\cite{Aad:2012dlq} and CMS\cite{Chatrchyan:2013yna}; (2) Fractions of $\Upsilon(nS)$ production originating from $\chi_{b}(nP)(n=1,2,3)$ freedown contributions measured by LHCb~\cite{Aaij:2014caa} which are denoted as $R_{\Upsilon(mS)}^{\chi_b(nP)}$ (values for $m\neq n$ are not included in the fit but predicted by using the branching ratios in TABLEs 1 and 2 and compared with data, as shown in Fig.2); (3) Cross section ratio of $\chi_{b2}(1P)$ to $\chi_{b1}(1P)$ measured by CMS~\cite{Khachatryan:2014ofa}. To avoid potential non-perturbative effects in the sense that only the first two powers in the $1/p_T^2$ expansion of cross sections are proven to be factorizable \cite{Kang:2014tta}, we need to introduce a relatively large $p_T$ cutoff for the data (for the similar case in the production of $\psi^{(\prime)}$, see Refs.~\cite{Ma:2010yw,Ma:2010jj,Faccioli:2014cqa}). In our fit, we only use data in the region $p_T>15$ GeV because the $\chi^2/d.o.f.$ will increase quickly when the $p_T$ cutoff becomes smaller than $15$ GeV. For example, by choosing the $p_T$ cutoff to be  $7, 9, 11, 13, 15$, and $17$ GeV,  the corresponding $\chi^2/d.o.f.$ in fitting $\Upsilon(3S)$ data are $4.2, 4.0, 2.5, 1.9, 1.3$, and $1.0$, respectively.

\begin{figure*}[!hbtp]
\begin{center}
\includegraphics*[scale=0.42]{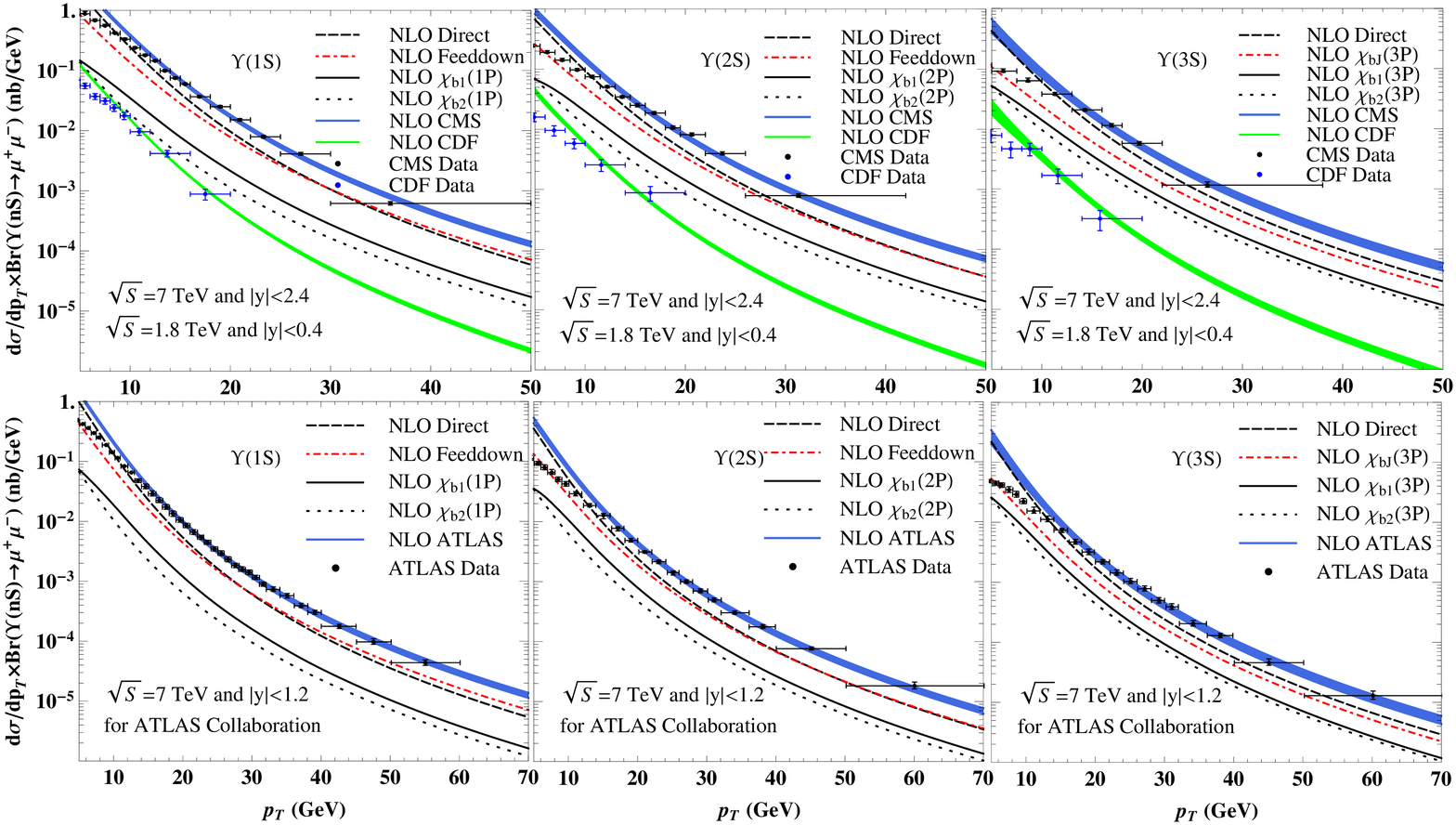}
\caption{Differential $p_T$ cross sections for the experimental windows of ATLAS, CMS and CDF. From left to right: $\Upsilon(1S)$, $\Upsilon(2S)$, $\Upsilon(3S)$. The contributions from direct production are denoted by dashed lines, while those from feeddown by dashed-dotted lines. The $\chi_{b1}(nP)-\Upsilon(nS)$ and $\chi_{b2}(nP)-\Upsilon(nS)$ feeddown contributions are denoted by the solid and dotted lines, respectively. The experimental data are taken from Refs.~\cite{Aad:2012dlq,Chatrchyan:2013yna,Acosta:2001gv}.}
 \label{fig-1}
 \end{center}
\end{figure*}

\begin{figure*}[!hbtp]
\begin{center}
\includegraphics*[scale=0.42]{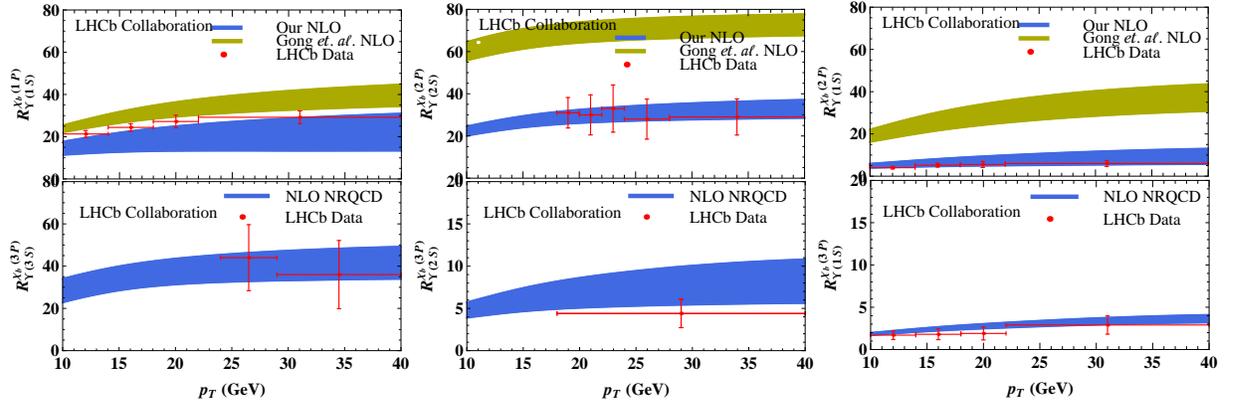}
\caption{The fractions of $\Upsilon(mS)~(m=1,2,3)$ production originating from $\chi_b(nP)~(n=1,2,3;~n\geq m)$ feeddown contributions, denoted as $R_{\Upsilon(mS)}^{\chi_b(nP)}$ (in units of percentage). From left to right: $R_{\Upsilon(1S)}^{\chi_b(1P)}$, $R_{\Upsilon(2S)}^{\chi_b(2P)}$, $R_{\Upsilon(1S)}^{\chi_b(2P)}$ in the first row and $R_{\Upsilon(3S)}^{\chi_b(3P)}$, $R_{\Upsilon(2S)}^{\chi_b(3P)}$, $R_{\Upsilon(1S)}^{\chi_b(3P)}$ in the second row. Our predictions are denoted by the blue bands, while those obtained by using parameters in Ref.\cite{Gong:2013qka} are denoted by the yellow bands. Experimental data are taken from Ref.\cite{Aaij:2014caa}.}
 \label{fig-4}
 \end{center}
\end{figure*}

When $p_T>15$ GeV, we find the CO P-wave  $^3\!P_J^{[8]}$ contribution can be decomposed into a linear combination of $^1\!S_0^{[8]}$ and $^3\!S_1^{[8]}$ (just similar to the $J/\psi$ case~\cite{Ma:2010yw,Ma:2010jj}),
\begin{eqnarray}
\label{deco}
{\rm d}\hat{\s}(^3\!P_J^{[8]})=r_0~{\rm d}\hat{\s}(^1\!S_0^{[8]})+r_1~{\rm d}\hat{\s}(^3\!S_1^{[8]}),
\end{eqnarray}
where $r_0=3.8$, $r_1=-0.52$, which may slightly change with rapidity ranges. So with three CO LDMEs we can extract two linear combinations, which are denoted by
\begin{eqnarray}
\label{M0M1}
M_{0,r_0}^{\Upsilon(nS)}=\langle\mathcal{O}^{\Upsilon(nS)}(^1\!S_0^{[8]})\rangle+\frac{r_0}{m_b^2}\langle\mathcal{O}^{\Upsilon(nS)}(^3\!P_0^{[8]})\rangle,
\\ \nonumber
M_{1,r_1}^{\Upsilon(nS)}=\langle\mathcal{O}^{\Upsilon(nS)}(^3\!S_1^{[8]})\rangle+\frac{r_1}{m_b^2}\langle\mathcal{O}^{\Upsilon(nS)}(^3\!P_0^{[8]})\rangle,
\end{eqnarray}
which account for $1/p_T^6$ and $1/p_T^4$ behaviors, respectively.

\begin{table}[!htb]
\begin{tabular}{{c}c*{3}{c}}
\hline\hline \itshape ~~~~~&~$\langle\mathcal{O}(\ss)\rangle$~ &
~$M_{0,r_0}$~&
~$M_{1,r_1}$~&
\\
\itshape  ~~~~ & $\rm{GeV}^3$ & $10^{-2}\rm{GeV}^3$ & $10^{-2}\rm{GeV}^3$
\\\hline $\Upsilon(1S)$ &~9.28~& $13.70\pm1.11$& $1.17\pm0.02$
\\\hline $\Upsilon(2S)$ &~4.63~& $6.07\pm1.08$& $1.08\pm0.20$
\\\hline $\Upsilon(3S)$ &~3.54~& $2.83\pm0.07$& $0.83\pm0.02$
\\\hline\hline
\end{tabular}
\caption{\label{ns} The LDMEs for $\Upsilon(1S,2S,3S)$ production. The combined LDMEs are obtained by the fit, while the CS ones are estimated by using the $B-T$ potential model in Ref.\cite{Eichten:1995ch}.}
\end{table}

Based on the above method, we fit two linear combinations $M_{0,r_0}^{\Upsilon(nS)}$ and $M_{1,r_1}^{\Upsilon(nS)}$ for $\Upsilon(1S,2S,3S)$ with $\chi^2/\text{d.o.f}= 0.99, 2.07, 1.25$, together with CS LDMEs that are estimated by using the B-T potential model~\cite{Eichten:1995ch}(see Tab.\ref{ns}). As for $r_{nP}$, the results are listed in Tab.\ref{rnp-fit}, with those obtained in Ref.~\cite{Gong:2013qka} for comparison.
\begin{table}[!htp]
\begin{tabular}{{{c}ccc}}
\hline\hline \itshape
~$r_{nP}$~&~$n = 1$~&~$n = 2$~&~$n = 3$~
\\\hline This work & $0.42\pm0.05$ & $0.62\pm0.08$ & $0.83\pm0.22$
\\\hline Ref.~\cite{Gong:2013qka} & $0.85\pm0.11$ & $1.58\pm0.38$ &
\\\hline\hline
\end{tabular}
\caption{\label{rnp-fit} The values of $r_{nP}$ for $n=1,2,3$ in this work and in Ref.~\cite{Gong:2013qka}.}
\end{table}
In Tab.~\ref{ns},
we find that the central value of $M_{0,r_0}^{\Upsilon(nS)}$ decreases more quickly than that of $M_{1,r_1}^{\Upsilon(nS)}$ as $n$ increases, while the values of $M_{1,r_1}^{\Upsilon(nS)}$ almost have no changes. This explains why a higher $\Upsilon(nS)$ tends to have a less steep $p_T$ cross sections.

\begin{figure}
\includegraphics[width=7cm]{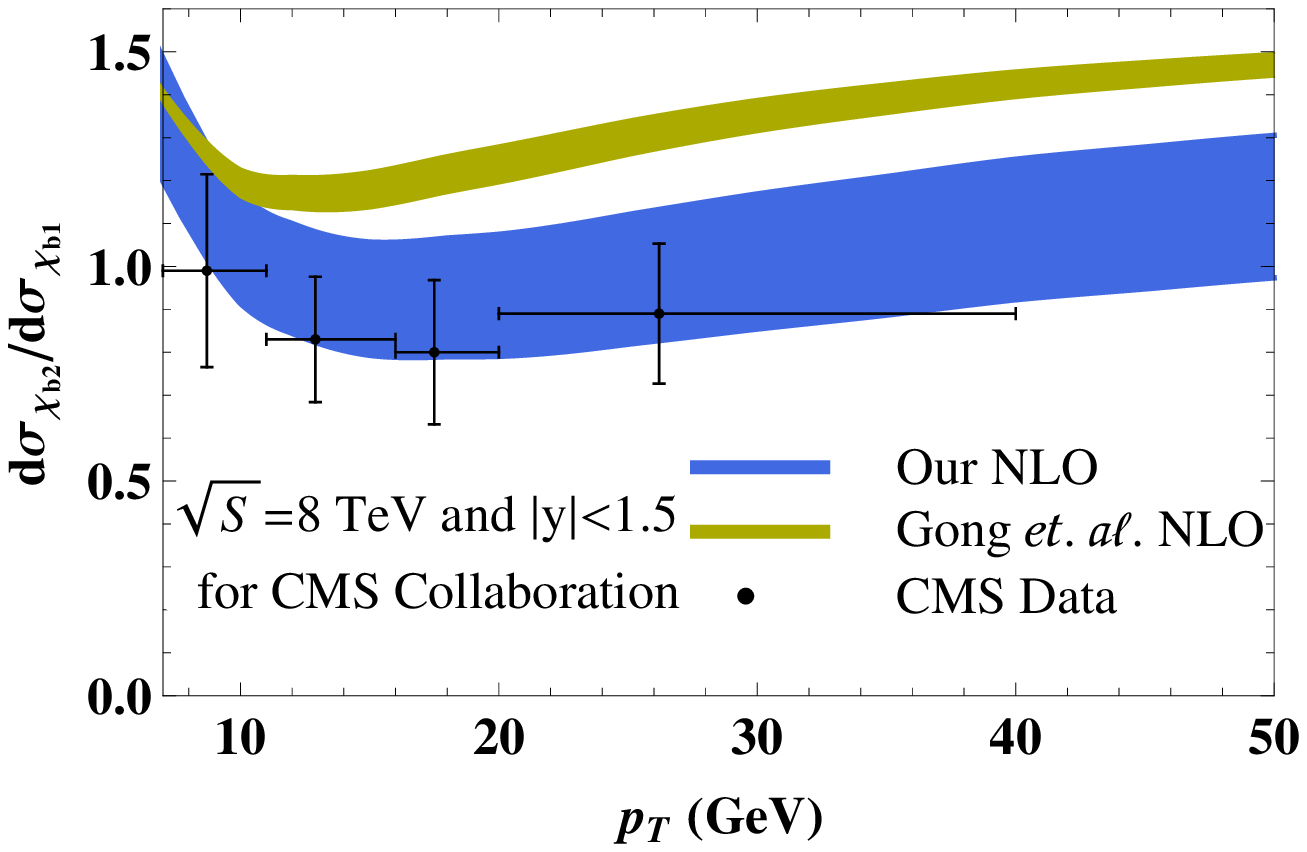}
\caption{The ratio of differential cross sections of $\chi_{b2}(1P)$ to $\chi_{b1}(1P)$  for the experimental windows of CMS. The blue band is our NLO results with the extracted value of $r_{1P}$ in Tab.~\ref{rnp-fit} and the yellow band is obtained by using parameters in Ref.\cite{Gong:2013qka}. Experimental data are taken from Ref.\cite{Khachatryan:2014ofa}}.
 \label{fig-3}
\end{figure}

Comparisons between our fit and data are shown in Figs.~\ref{fig-1}, \ref{fig-4} and \ref{fig-3}, along with our postdiction for the CDF cross section~\cite{Acosta:2001gv}. It is interesting to see that the yield, fractions of $\Upsilon(mS)$ production from $\chi_{b}(nP)$ decays, and cross section ratios for $\Upsilon(1S,2S,3S)$ can be well described simultaneously. In particular, good agreement with $R_{\Upsilon(nS)}^{\chi_b(3P)}$ is achieved explicitly by a relatively large feeddown contribution from $\chi_b(3P)$, as indicated by the large value of $r_{3P}$ in Tab.~\ref{rnp-fit}. For comparison, we also present the fractions $R_{\Upsilon(mS)}^{\chi_b(nP)}$ using the parameters in Ref.~\cite{Gong:2013qka}, which are shown in Fig.~\ref{fig-4} as the yellow bands. From Fig.~\ref{fig-4}, one sees that the $\chi_b(1P,2P)$ production rates predicted by Ref.~\cite{Gong:2013qka} are too large compared with data, whereas our predictions of the production rates of $\chi_b(1P,2P)$  and $\chi_b(3P)$, denoted by the blue bands in Fig.~\ref{fig-4}, are roughly consistent with data. In Fig.~\ref{fig-3}, with the extracted value of $r_{1P}$ in Tab.~\ref{rnp-fit} we can well describe the measured ratio of differential cross sections of $\chi_{b2}$ to $\chi_{b1}$ by CMS~\cite{Khachatryan:2014ofa}, clearly better than that in Ref.\cite{Gong:2013qka}.

\begin{figure*}[!hbtp]
\begin{center}
\includegraphics*[scale=0.39]{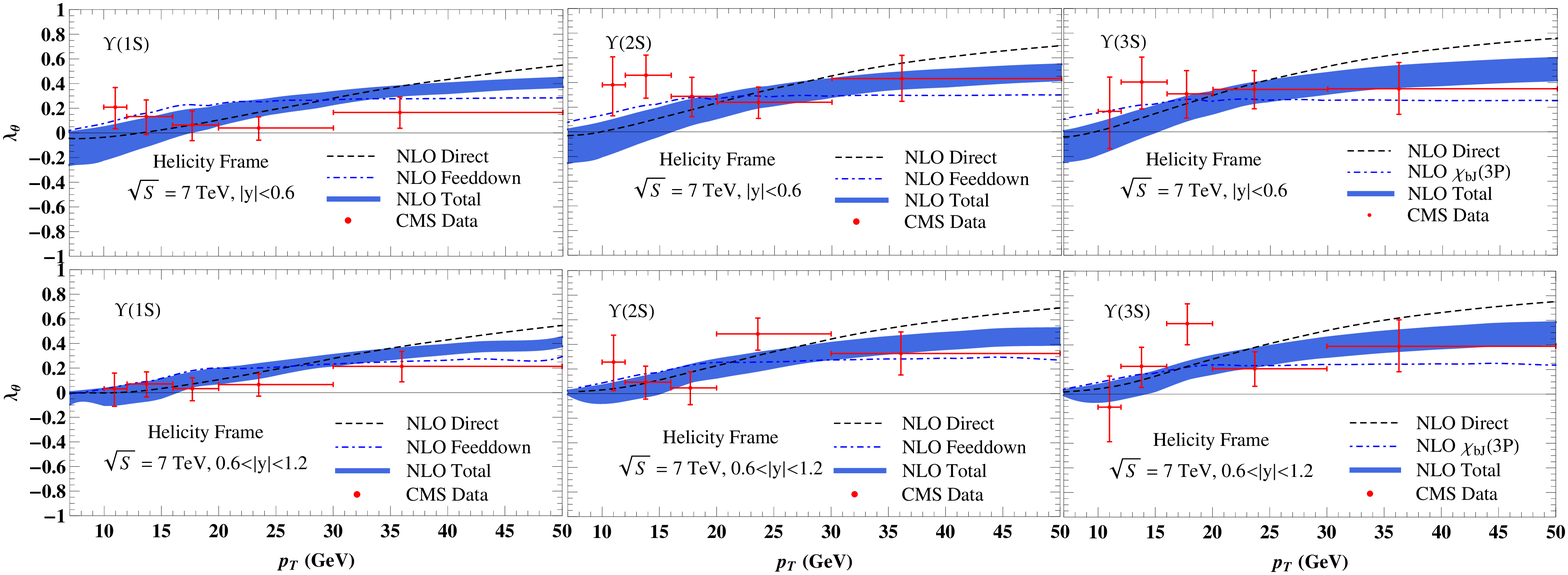}
\caption{The polarization parameter $\lambda_\theta$ in the helicity frame for the experimental widows at the LHC. From left to right: $\Upsilon(1S)$, $\Upsilon(2S)$, $\Upsilon(3S)$. The contributions from direct production are denoted by dashed lines, while those from feeddown by dashed-dotted lines. The total results are denoted by the blue bands. The experimental data are taken from Ref.\cite{Chatrchyan:2012woa}.}
 \label{fig-2}
 \end{center}
\end{figure*}


With the LDMEs extracted from yield data, we can calculate the $\Upsilon(nS)$ polarizations. The predicted $\lambda_\theta$ of $\Upsilon(1S,2S,3S)$ are the weighted averages of the direct production and feeddown contributions. This can be seen directly from Fig.~\ref{fig-2}, where the results for the CMS window at $\sqrt{S}=7$~GeV are shown.  The predictions for prompt $\Upsilon(1S,2S,3S)$ polarizations are roughly consistent with data. Note that the $\Upsilon(3S)$ polarization is obtained with a relatively large feeddown contribution from
$\chi_b(3P)$ (see the feeddown fraction $R_{\Upsilon(3S)}^{\chi_b(3P)}$ shown in Fig.~\ref{fig-4}),
which reduces the value of $\lambda_\theta$ of direct production and leads to a smaller total polarization $\lambda_\theta$ of prompt $\Upsilon(3S)$. The feeddown contributions also affect the $\Upsilon(1S,2S)$ polarizations and lead to better agreement with data.

In fact, the predicted $\lambda_\theta$'s of the prompt $\Upsilon(1,2,3S)$ are the weighted averages of the contributions from direct production and feeddown processes. This can be seen from Fig.~\ref{fig-2}. In particular, for the $\lambda_\theta$ of $\Upsilon(3S)$, the weight of feeddown contribution is just the fraction $R_{\Upsilon(3S)}^{\chi_b(3P)}$ shown in Fig.~\ref{fig-4}, which is as large as about $40\%$, as observed by LHCb~\cite{Aaij:2014caa}.
Since the fraction $R_{\Upsilon(3S)}^{\chi_b(3P)}$ is determined by the product of the $\chi_b(3P)$ production cross section and the branching ratio of $\chi_b(3P)\rightarrow \Upsilon(3S)\gamma$, a change of the branching ratio will cause a change of $\chi_b(3P)$ production cross section but keep the fitted fraction $R_{\Upsilon(3S)}^{\chi_b(3P)}$ unchanged. Namely, the uncertainty in the predicted branching ratio in Tab.~\ref{br} will affect the predicted value of $\chi_b(3P)$ cross section but not $R_{\Upsilon(3S)}^{\chi_b(3P)}$. As a result, the predicted polarization value $\lambda_{\theta}$ of the prompt $\Upsilon(3S)$ is insensitive to the input  branching ratio of $\chi_b(3P)\rightarrow \Upsilon(3S)\gamma$ but sensitive to the observed feeddown fraction $R_{\Upsilon(3S)}^{\chi_b(3P)}$.

\section{Summary}

At NLO in NRQCD, we study the $\Upsilon(nS)$ and $\chi_b(nP)$ (n=1,2,3) production at the LHC. We extract the LDMEs of $\Upsilon(nS)$ and $\chi_b(nP)$ production from the LHC large $p_T$ yield data \cite{Aad:2012dlq,Chatrchyan:2013yna,Khachatryan:2014ofa,Aaij:2014caa}, and then with these LDMEs make predictions for the $\Upsilon(nS)$ polarizations. We find that for large $p_T$ ($>$15~GeV) while the observed $\Upsilon(nS)$ differential $p_T$ cross sections, the fractions of $\Upsilon(mS)$ production from $\chi_b(nP)$ decays, and the differential cross section ratio of $\chi_{b2}(1P)$ to $\chi_{b1}(1P)$) can be rather well described, the predicted  $\Upsilon(1S,2S,3S)$ polarizations also agree with the recent measurements by CMS \cite{Chatrchyan:2012woa} within errors. As a result, a simultaneously good description for the large $p_T$ cross sections and polarizations of $\Upsilon(1S,2S,3S)$ is achieved at NLO in NRQCD. In particular, the prompt $\Upsilon(3S)$ polarization puzzle can be understood with a large feeddown contribution from $\chi_b(3P)$ states.

\begin{acknowledgments}

We thank S.~Argiro, V.~Belyaev, and Z.~Yang for useful discussions on quarkonia experiments at the LHC. This work was supported by the National Natural
Science Foundation of China (No. 11475005, No.11075002, No.11021092), and the National Key Basic Research Program of China (No 2015CB856700). Y.Q.M. is supported by the U.S. Department of Energy Office of Science, Office of Nuclear Physics under Grant
No.DE-FG02-93ER-40762.

\end{acknowledgments}




\begin{thebibliography}{10}



\bibitem{Abe:1992ww}
  F.~Abe {\it et al.}  [CDF Collaboration],
  Phys.\ Rev.\ Lett.\  {\bf 69}, 3704 (1992).



\bibitem{Braaten:1994vv}
  E.~Braaten and S.~Fleming,
  Phys.\ Rev.\ Lett.\  {\bf 74}, 3327 (1995).


\bibitem{Bodwin:1994jh}
  G.~T.~Bodwin, E.~Braaten and G.~P.~Lepage,
  Phys.\ Rev.\ D {\bf 51}, 1125 (1995);
  Phys.\ Rev.\ D {\bf 55}, 5853 (1997).


\bibitem{Brambilla:2010cs}
  N.~Brambilla {\it et al.},
  Eur.\ Phys.\ J.\ C {\bf 71}, 1534 (2011).



\bibitem{Butenschoen:2012px}
  M.~Butenschoen and B.~A.~Kniehl,
  Phys.\ Rev.\ Lett.\  {\bf 108}, 172002 (2012).


\bibitem{Chao:2012iv}
  K.~T.~Chao, Y.~Q.~Ma, H.~S.~Shao, K.~Wang and Y.~J.~Zhang,
  Phys.\ Rev.\ Lett.\  {\bf 108}, 242004 (2012).


\bibitem{Gong:2012ug}
  B.~Gong, L.~P.~Wan, J.~X.~Wang and H.~F.~Zhang,
  Phys.\ Rev.\ Lett.\  {\bf 110}, 042002 (2013).

\bibitem{Bodwin:2014gia}
  G.~T.~Bodwin, H.~S.~Chung, U.~R.~Kim and J.~Lee,
  Phys.\ Rev.\ Lett.\  {\bf 113}, 022001 (2014).


\bibitem{Chatrchyan:2012woa}
  S.~Chatrchyan {\it et al.}  [CMS Collaboration],
  Phys.\ Rev.\ Lett.\  {\bf 110}, 081802 (2013).

\bibitem{Braaten:2000gw}
  E.~Braaten and J.~Lee,
  Phys.\ Rev.\ D {\bf 63}, 071501 (2001).

\bibitem{Artoisenet:2008fc}
  P.~Artoisenet, J.~M.~Campbell, J.~P.~Lansberg, F.~Maltoni and F.~Tramontano,
  Phys.\ Rev.\ Lett.\  {\bf 101}, 152001 (2008).


\bibitem{Wang:2012is}
  K.~Wang, Y.~Q.~Ma and K.~T.~Chao,
  Phys.\ Rev.\ D {\bf 85}, 114003 (2012).

\bibitem{Likhoded:2012hw}
  A.~K.~Likhoded, A.~V.~Luchinsky and S.~V.~Poslavsky,
  Phys.\ Rev.\ D {\bf 86}, 074027 (2012)

\bibitem{Gong:2013qka}
  B.~Gong, L.~P.~Wan, J.~X.~Wang and H.~F.~Zhang,
  Phys.\ Rev.\ Lett.\  {\bf 112}, 032001 (2014).



\bibitem{Khachatryan:2014ofa}
  V.~Khachatryan {\it et al.}  [CMS Collaboration],
  Phys.\ Lett.\ B {\bf 743}, 383 (2015).


\bibitem{Aaij:2014caa}
  R.~Aaij {\it et al.}  [LHCb Collaboration],
  Eur.\ Phys.\ J.\ C {\bf 74}, 3092 (2014).


\bibitem{Ma:2010yw}
  Y.~Q.~Ma, K.~Wang and K.~T.~Chao,
  Phys.\ Rev.\ Lett.\  {\bf 106}, 042002 (2011).

\bibitem{Ma:2010jj}
  Y.~Q.~Ma, K.~Wang and K.~T.~Chao,
  Phys.\ Rev.\ D {\bf 84}, 114001 (2011).


\bibitem{Shao:2012iz}
  H.~S.~Shao,
  Comput.\ Phys.\ Commun.\  {\bf 184}, 2562 (2013).


\bibitem{Whalley:2005nh}
  M.~R.~Whalley, D.~Bourilkov and R.~C.~Group,
  hep-ph/0508110.


\bibitem{Shao:2012fs}
  H.S. Shao and K.T. Chao,
  Phys.\ Rev.\ D {\bf 90}, 014002 (2014).

\bibitem{Shao:2014fca}
  H.~S.~Shao, Y.~Q.~Ma, K.~Wang and K.~T.~Chao,
  Phys.\ Rev.\ Lett.\  {\bf 112}, 182003 (2014).
  
\bibitem{Ma:2010vd}
  Y.~Q.~Ma, K.~Wang and K.~T.~Chao,
  Phys.\ Rev.\ D {\bf 83}, 111503 (2011).


\bibitem{Eichten:1995ch}
  E.J.~Eichten and C.~Quigg,
  Phys.\ Rev.\ D {\bf 52},1726(1995).


\bibitem{Kwong:1988ae}
  W.~Kwong and J.L.~Rosner,
  Phys.\ Rev.\ D {\bf 38},279(1988).

\bibitem{Godfrey:2015dia}
  S.~Godfrey and K.~Moats,
  Phys.\ Rev.\ D {\bf 92}, no. 5, 054034 (2015)

\bibitem{Beringer:1900zz}
  J.~Beringer {\it et al.}  [Particle Data Group],
  Phys.\ Rev.\ D {\bf 86}, 010001 (2012).



\bibitem{Aad:2012dlq}
  G.~Aad {\it et al.}  [ATLAS Collaboration],
  Phys.\ Rev.\ D {\bf 87}, 052004 (2013).


\bibitem{Chatrchyan:2013yna}
  S.~Chatrchyan {\it et al.}  [CMS Collaboration],
  Phys.\ Lett.\ B {\bf 727}, 101 (2013).

\bibitem{Kang:2014tta}
  Z.~B.~Kang, Y.~Q.~Ma, J.~W.~Qiu and G.~Sterman,
  Phys.\ Rev.\ D {\bf 90}, 034006 (2014).

\bibitem{Faccioli:2014cqa}
  P.~Faccioli, V.~Knunz, C.~Lourenco, J.~Seixas and H.~K.~Wohri,
  Phys.\ Lett.\ B {\bf 736}, 98 (2014).

\bibitem{Acosta:2001gv}
  D.~Acosta {\it et al.}  [CDF Collaboration],
  Phys.\ Rev.\ Lett.\  {\bf 88}, 161802 (2002).









\end{thebibliography}
\providecommand{\href}[2]{#2}\begingroup\raggedright
\end{document}